\documentclass[11pt]{article}   

\usepackage{amsmath,amssymb,graphicx}

\usepackage{bm}

\def\atan{\mathop{\rm arctan}\nolimits}
\def\ath{\mathop{\rm tanh^{-1}}\nolimits}
\def\const{\mathop{\rm const}\nolimits}
\def\n{\bgroup\bf n\egroup}
\def\m{\bgroup\bf m\egroup}
\def\x{\bgroup\bf x\egroup}
\def\uu{\bgroup\bf u\egroup}
\def\g{a^2\cos^2 v+c^2\sin^2 v}
\def\d{\sqrt{1-\eta^2}}

\begin{document}

\begin{center}
{\Large \bf Topological defects and shape of aromatic self-assembled vesicles}\\[2ex]
{\large O. V. Manyuhina, A. Fasolino and M. I. Katsnelson\\[1ex]
{\large \it Institute for Molecules and Materials, Radboud University Nijmegen, Heyendaalseweg 135, 6525~AJ Nijmegen, The Netherlands}}
\end{center}


\begin{abstract}
We show that the stacking of flat aromatic molecules on a curved surface results in topological defects. We consider, as an example,  spherical vesicles, self-assembled from molecules with 5- and 6-thiophene cores. We predict that the symmetry of the molecules influences the number of topological defects and the resulting equilibrium shape. 
\end{abstract}

Engineering of molecular assemblies is one of the aims of modern nanotechnology. Here, starting from ideas put forward by Nelson, MacKintosh and Lubensky for liquid crystals~\cite{nelson:2002,lubensky:1992,mackintosh:1991} we show that topological defects can determine the shape of supramolecular vesicles\cite{antonietti}. Recently, Nelson~\cite{nelson:2002} proposed an elegant way to generate topological defects by coating spherical colloidal particle  with anisotropic object like nematic liquid crystals. The high-energy core of defects can be used as sites for chemical activity and as potential spots for the assembly of three dimensional architectures. Further, the classical works~\cite{lubensky:1992,mackintosh:1991} relate the appearance of topological defects in lipid bilayers to the temperature-induced phase transition from the Sm-A phase to the Sm-C phase with tilted molecules.

In the last decades supramolecular chemistry has created large aromatic molecules that form a a variety of nanometer sized structures~\cite{rowan:2003,fuhrop:2004}. Here we demonstrate that packing of flat aromatic molecules on a curved surface, e.g. on a sphere, results in unavoidable topological defects, similar to those considered by Nelson. Unlike colloidal particles and two dimensional spherical crystals~\cite{bowick:2000,bowick:2003}, the defects in aromatic self-assemblies could influence the geometry and the topology of the structure as a whole. Therefore, we allow spherical vesicles to deform towards spheroids in order to determine the minimum energy configuration.

\begin{figure}[t]
\centering
\includegraphics[width=0.8\linewidth]{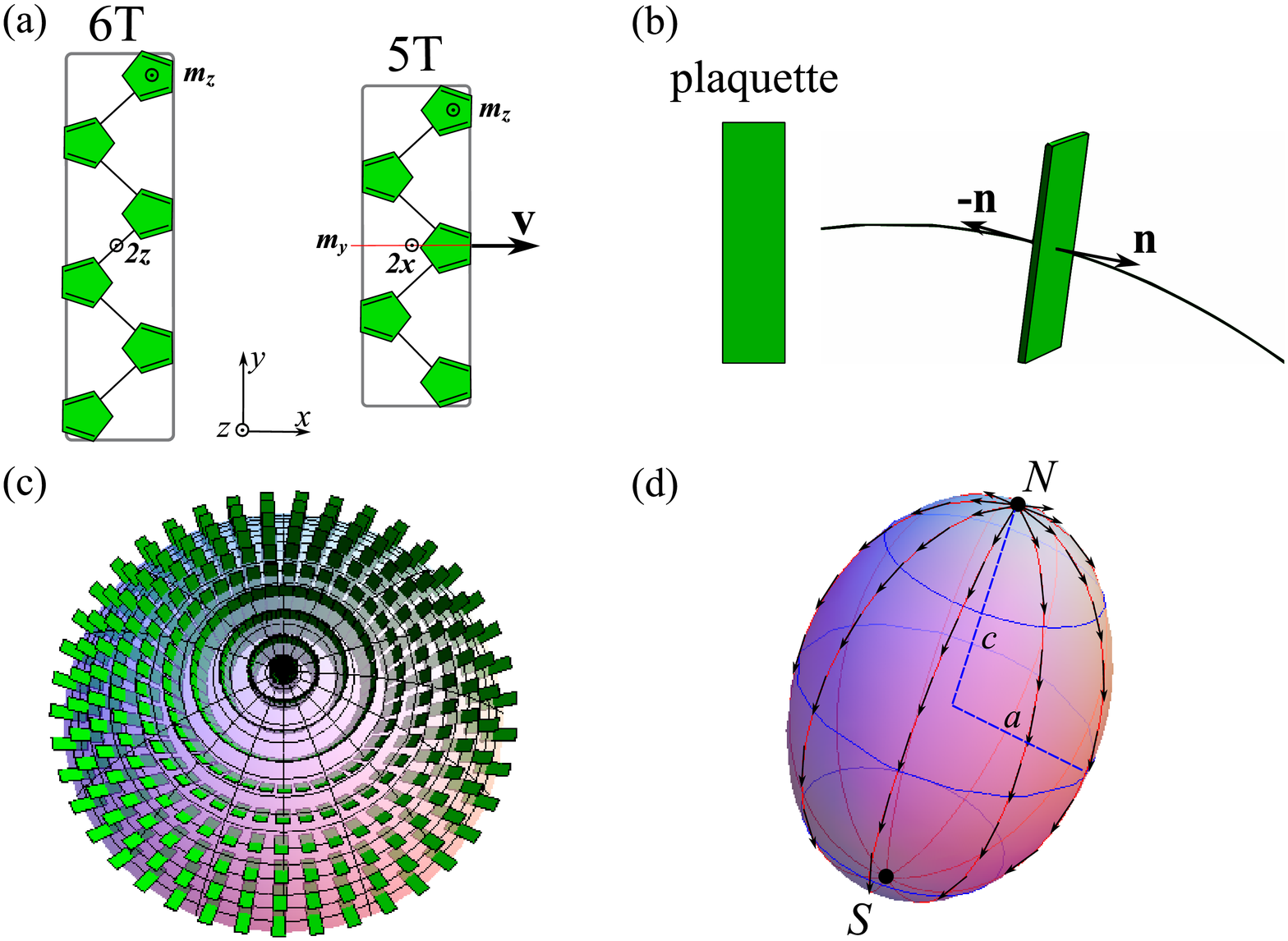}
\caption{(a) Aromatic molecules with six thiophene (6T) cores and five thiophene cores (5T) that can be thought of as flat plaquettes (b)~Since for 6T there is no invariant vector ${\bf v}$, the normal to the plane is a true vector {\bf n} in the $z$-direction. For 5T we can identify the invariant vector ${\bf v}$ (in $x$-direction), and thus there is a director ${\bf n}\to -{\bf n}$ in the $z$-direction. (c)~The arrangement of flat plaquettes on a spherical cap, resulting in a topological defect at the center.  (d) The tangent vector field and two topological defects (thick dots) located at the north ($N$) and the south ($S$) poles of a spheroid; $c$ and $a$ are the semiaxes of the spheroid.}
\label{fig1}
\end{figure}

As an example, consider the bolaamphiphilic sexithiophene (6T) molecules~\cite{fuhrop:2004,shklyarevskiy} known to self-assemble in propanol-2 into spherical vesicles with radius of the order of 100 nm. The aromatic 6T core  can be thought of as a flat plaquette\footnote{Bolaamphiphile 2,5'''''-(R-2-methyl-3,6,9,12,15-pentaoxahexadecyl ester) 6T  has two polar ethylene oxide chains, which do not play any role in our consideration.} (see Fig.~\ref{fig1}a,b). Consider now the regular arrangement of flat plaquettes on a curved surface, e.g. a spherical cap (see Fig.~\ref{fig1}c): every plaquette can be described by a director (headless vector) normal to the plane of the plaquette and thus tangent to a sphere. The hairy ball theorem (or generalized Poincare-Hopf theorem) states that there is no nonvanishing continuous tangent (vector) field on a sphere. This means that we cannot pack flat molecules on a sphere without creating a topological defect, namely, a point around which the tangent field rotates by $2\pi q$,  with $q$ an integer or half-integer number, like the one shown in Fig.~\ref{fig1}c. The number of topological defects depends on the Euler characteristic of the surface and on the intrinsic symmetry of the self-assembled molecules. The 6T core, enclosed in a rectangle,  has $2/m$ lattice symmetry\cite{crystallogr}, which means that there are no invariant vectors in the plane of the molecule or that there is inversion symmetry. This defines a  unique normal n to the plane of the plaquette which will be a tangent vector to a sphere. Instead, five-thiophene groups (5T)  have $mm2$ point group transformations, which leave a vector ${\bf v}=(x,0,0)$ invariant under symmetry operations. This additional symmetry of the core yields a director to the plane of the molecule with ${\bf n}\to-{\bf n}$. We may generalize our results to even and odd number of thiophene and benzene rings. For any aromatic molecule, it is always  possible to assign either a true vector normal to the plaquette or a director. For the tangent vector field on a sphere, there are two topological defects  with vorticity $2\pi$ (topological charge $q=+1$), whereas for directors there are four defects with  vorticity $\pi$ ($q=+1/2$)~\cite{nelson:2002,lubensky:1992}. Based on the example of 6T molecules, which can be generalized to any aromatic molecule with inversion symmetry, we consider in the following the interaction of two topological defects on surfaces with the topology of a sphere.

Like  charged particles, the defects interact with each other and with the Gaussian curvature as a background charge distribution~\cite{vitelli:2004,vitelli:2009}. Taking into account the fact that self-assembled structures  can adjust their shape in order to minimize the free energy, we allow deformations of the sphere towards spheroids. This feature is essentially different from the case of colloidal particles where the geometry is determined by the substrate~\cite{nelson:2002}. The quantitative understanding of the energetics of topological defects interacting on spheroids involves the calculation of the Green function $\Gamma({\bf x}_i,{\bf x}_j)$, which is the inverse of the Laplace--Beltrami operator on spheroids~\cite{nelson:2002, bowick:2000}. This is a non-trivial problem, which can be solved through the conformal mapping of the complex surface onto a simple one, e.g. plane or sphere, where the Green function is known~\cite{vitelli:2009}. 

To quantify the effect of topological defects  we calculate the total free energy $F_{\rm total}$ as the sum of bending energy $F_{\rm bend}$ and of the energy associated with topological defects $F_{\rm defect}$ on spheroids, namely 
\begin{equation}\label{eq1}
F_{\rm total} = F_{\rm bend} + F_{\rm defect} = 2k\iint \!dS\, H^2 + 4\pi^2 K_A q_iq_j\Gamma ({\bf x}_i,{\bf x}_j),
\end{equation}
where $H$ is the mean curvature, $k$ is the bending rigidity, and $K_A$ is hexatic constant~\cite{lubensky:1992,helfrich:1973}. For every point of the surface one can define the two principal curvatures $\kappa_1$ and $\kappa_2$. The mean curvature $H=(\kappa_1+\kappa_2)/2$ at a given point depends on the change of the normal vector around this point, characterizing the local properties of the surface. The Gaussian curvature $K=\kappa_1\kappa_2$, instead, is an intrinsic property of the surface defined only by its metric, and its integral over a smooth surface $\int dS\,K$ is related by Gauss--Bonnet theorem~\cite{kamien:2002} to the Euler characteristic $\chi$, which determines the topology of the surface. Since we consider spheroids with $\chi=2$, the integral over $K$ is constant and therefore it is not included in Eq.~(\ref{eq1}). The non-local (long-ranged) interactions of the defects are described by the Green function $\Gamma({\bf x}_i,{\bf x}_j)$, which depends only on the metric of the surface. The competition between two terms in Eq.~(\ref{eq1}) leads to frustration, because the defects favour high Gaussian curvature~\cite{vitelli:2004}, which inevitably leads to high mean curvature (due to $H^2\geqslant K$), and consequently to an increase of the bending energy $F_{\rm bend}$. Because of the repulsive nature of defect interactions, the minimum of the second term corresponds to defects located at the north ($N$) and the south ($S$) poles of a spheroid for a tangent vector field (see Fig.~\ref{fig1}d), and at the vertices of a tetrahedron for a director field~\cite{lubensky:1992}. The topological defects in self-assemblies account for the  competition between local interactions of aromatic cores and the topological constraints, whereas the details of microscopic potentials reside in the bending rigidity $k$ and the hexatic constant $K_A$. In our system, the ratio $K_A/k$ is the only governing parameter, related to details of  intermolecular interactions for a given temperature. Although the value of the bending rigidity $k=(2.56\pm 0.8)\cdot10^{-21}$~J was found from experimental measurements of magnetic deformations of 6T spherical vesicles in isopropanol~\cite{prl:2007}, the value of the hexatic constant $K_A$ is a less accessible parameter. Assuming that $\pi$-$\pi$ interactions between thiophene molecules give the main contribution to the values of $k$ and $K_A$ we can roughly estimate the ratio $K_A/k$ based on the results of {\it ab initio} molecular orbitals calculations for different geometries of thiophene dimer~\cite{tsuzuki:2002, marzari:2004}, yielding $K_A/k$ of the order of unity.

\begin{figure}[t]
\centering
\begin{minipage}[c][82mm][c]{0.75\linewidth}
\raisebox{82mm}{(a)}\includegraphics[width=\linewidth]{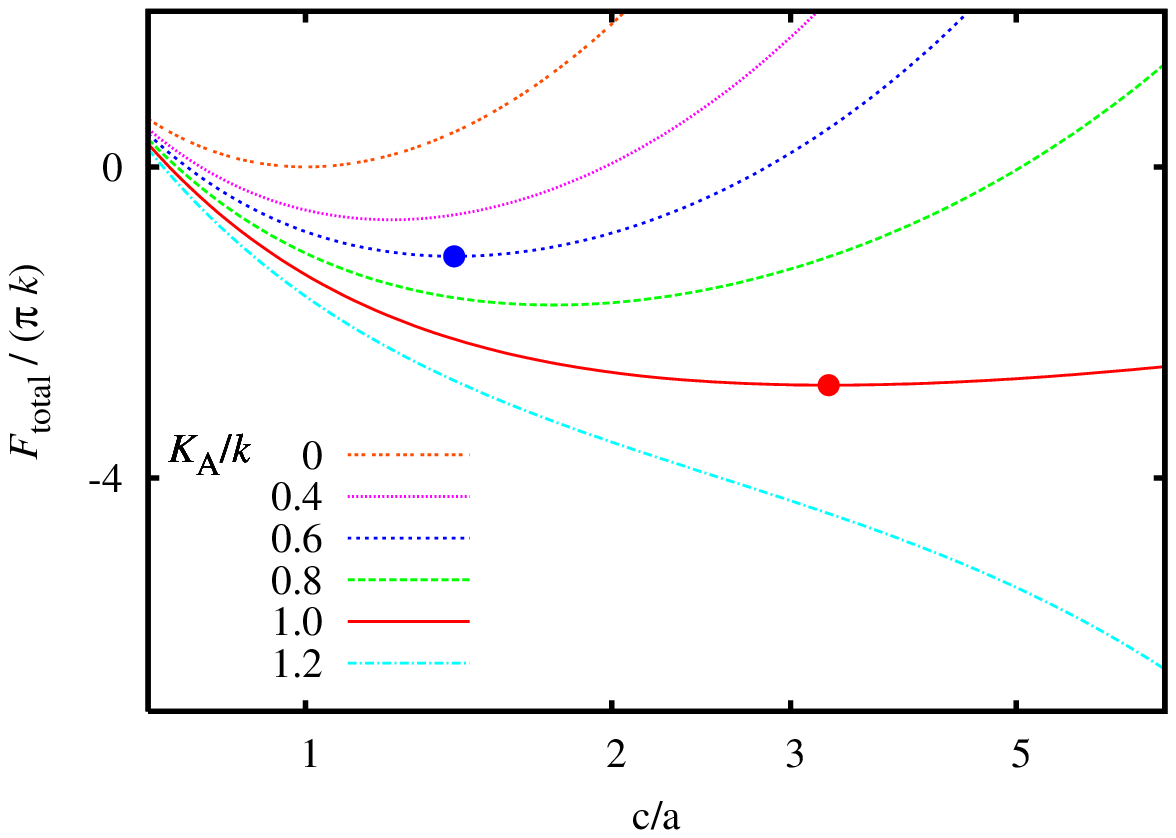}
\end{minipage}
\hfil
\raisebox{40mm}{(b)}\begin{minipage}[c][82 mm][c]{0.15\linewidth}
\centering
\includegraphics[scale=0.37]{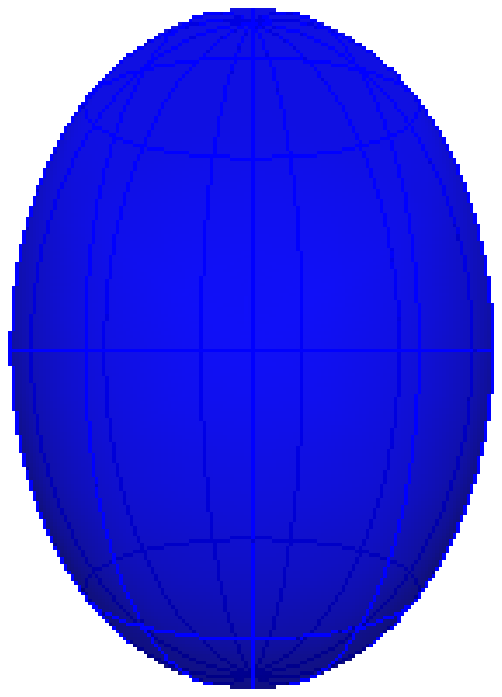}\\[2ex]
\includegraphics[scale=0.37]{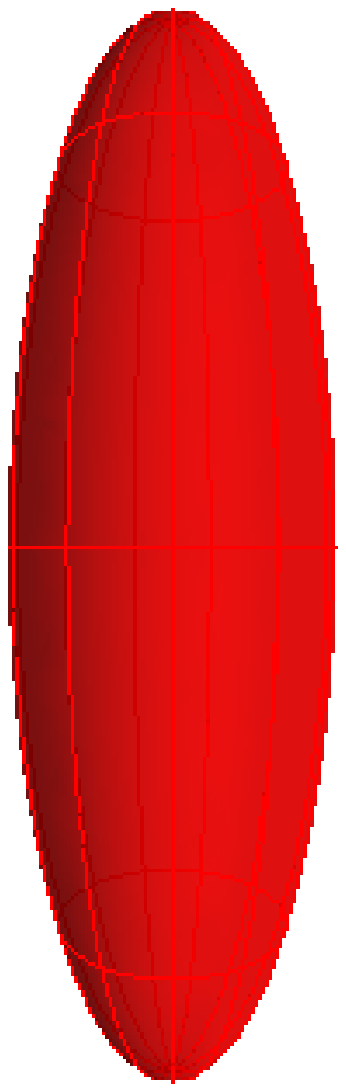}
\end{minipage}
\medskip
\caption{(a) The normalized free energy $F_{\rm total}/(\pi k)$ as function of $c/a$ (the axes of the spheroid, see Fig.~\ref{fig1}d) in logarithmic $x$-scale. The surface area of the vesicle is kept constant, assuming constant density of the molecules. The minimum of $F_{\rm total}$, corresponding to a sphere $c=a$ for $K_A=0$, shifts towards higher values of $c/a$  with hexatic constant $K_A$ compared to $k$. For the ratio $K_A/k = 1.2$ we expect an instability towards rod-like structures.  (b) The equilibrium shapes  marked with the thick dots on the plot.}
\label{fig2}
\end{figure}

It turns out, that for the case of spheroid the expression for the total free energy $F_{\rm total}$ (Eq.~(\ref{eq1})) can be given in simple analytical form as the function of $c/a$, the ratio of the semiaxes of spheroids, with bending energy
\begin{equation}\label{eq2}
F_{\rm bend} = 2k\pi (c/a)^2\bigg\{\frac{\tanh^{-1} \sqrt{1-(c/a)^2} }{\sqrt{1-(c/a)^2}}+\frac2{(c/a)^2}+\frac1{3(c/a)^2}\bigg(1+\frac2{(c/a)^2}\bigg)\bigg\},
\end{equation}
and the Green function
\begin{equation}\label{eq3}
\Gamma (0,\pi) = -\frac1{2\pi}\bigg(\log\frac{2a}d + \sqrt{-1+(c/a)^2} \tan^{-1}\sqrt{-1+(c/a)^2}\bigg).
\end{equation}
Note that since the Gaussian curvature at the poles is given by $K=c^2/a^4$, $\Gamma (0,\pi)$ becomes more negative for increasing $K$.
We plot in Fig.~\ref{fig2}a the total free energy $F_{\rm total}$ for different values of $K_A/k$ as a function of $c/a$ with defects located at $z=c$ ($N$) and $z=-c$ ($S$). In presence of topological defects, the equilibrium shapes, corresponding to the minimum of $F_{\rm total}$, are shown for two values of  $K_A/k$ (see Fig.~\ref{fig2}b). When the contribution $F_{\rm defect}$ is not dominant ($K_A\lesssim0.6k$) the equilibrium shapes are close to spherical, but for $K_A\simeq k$ they are elongated with $c/a\simeq 3.2$. For small values of the bending rigidity ($K_A>k$) we expect an instability towards infinitely long tubular structures. A similar elongation of phospholipid vesicles has  been predicted to occur as a function of temperature at the phase transition between Sm-A phase and Sm-C phase~\cite{mackintosh:1991}. Indeed, 6T molecules in butanol, in contrast with isopropanol, were found to self-assemble into multiwalled cylindrycal structures~\cite{gielen:2009}. The theory presented here holds only for single layered structures and thus we cannot establish a direct comparison with the experimental observations. Moreover, in real situations, parameters such as $k$ depend also on the particular solvent~\cite{fasolino:2006}.

In conclusion, we demonstrated that the stacking of flat aromatic molecules in self-assembled spherical vesicles leads to topological defects. The essential difference in symmetry between molecules with even and odd number of thiophene groups results in two and four topological defects, respectively.   In the case of two interacting defects the equilibrium shape of vesicles turns out to be elongated. We believe that four defects at the vertices of a tetrahedron would hardly lead to an elongation, although a detailed analysis of four interacting defects is beyond the scope of this paper.  In general, the question of how the  shape of self-assembled structures is connected with the  intrinsic properties, such as the symmetry, of the constituent molecules remains open. Here we proposed a simple geometric approach based on topological defects to establish this connection for  molecules with thiophene/benzene aromatic cores. The proposed theory can be used to extract the value of $K_A/k$ from experimental observations of the shape of self-assembled aromatic molecules.

\subsection*{Acknowledgements}
We thank Aloysio Janner for  his invaluable help in understanding the symmetry groups of aromatic molecules and Peter Christianen for fruitful discussions.

\subsection*{Supporting information}
Detailed analysis of the point group symmetries of 6T and 5T molecules, conformal mapping of spheroids, calculation of the Green function, derivation of the  analytical expression of $F_{\rm total}$ and estimate of $K_A/k$. This material is available free of charge via the Internet at http://pubs.acs.org.

\clearpage

\begin{center}
{\Large \bf
Supporting information for: Topological defects and shape of aromatic self-assembled vesicles}
\end{center}

\bigskip

\section{Point groups for 6T and 5T}

In Figure~1a we show two aromatic molecules (6T and 5T) with even and odd number of thiophene cores and the corresponding symmetry operations, which  transform the molecule into itself. For two-dimensional rectangular lattice, like the one enclosing 6T and 5T molecules, the possible symmetry  operations are mirror-reflections across line, denoted by $m$. For 6T molecule, we can write the following symmetry operations which leave one point unmoved~\cite{crystallogr}
\begin{equation}
m_z=\begin{pmatrix}1&&\\&1&\\&&-1\end{pmatrix} \quad\mbox{and}\quad 
2_z=m_xm_y=\begin{pmatrix}-1&&\\&-1&\\&&1\end{pmatrix}.
\end{equation}
Their product $m_xm_ym_z=\begin{pmatrix}-1&&\\&-1&\\&&-1\end{pmatrix}$, acting on a vector, results in the total inversion symmetry.  For 5T molecule the point group is different, containing the operators
\begin{equation}
m_z=\begin{pmatrix}1&&\\&1&\\&&-1\end{pmatrix}, \quad
m_y=\begin{pmatrix}1&&\\&-1&\\&&1\end{pmatrix}, \quad
2_x=m_ym_z=\begin{pmatrix}1&&\\&-1&\\&&-1\end{pmatrix}.
\end{equation}
These symmetry operations leave vector $v=(x,0,0)$ invariant, which is not the case for 6T molecules. This result can be generalized to even and odd number of thiophene and benzene rings.

\section{Conformal mapping}

Let us consider a spheroid, which is a surface of revolution  
given by the following parametrization
\begin{equation}\label{eqA:shape}
\x(u,v) = \big( a\sin v\cos u,\,a\sin v\sin u,\,c\cos v\big),
\qquad 0\leq v\leq\pi,\quad 0\leq u \leq 2\pi,
\end{equation}
with semiaxes $c$ and $a$ (see Figure~1d). Then, the metric of spheroid is given by
\begin{equation}\label{eq:spheroid}
ds_1^2 = (a^2\cos^2\! v+c^2\sin^2\!v)\,dv^2 + a^2\sin^2\! v\, du ^2.
\end{equation}
The mapping of spheroids onto a sphere with metric $ds^2 = R^2(d\theta^2 +\sin^2\!\theta\,d\phi^2)$ is said to be conformal if we can write
\begin{equation}\label{eq:map}
ds^2=e^{2\lambda(\uu)} ds_1^2,
\end{equation}
where $e^{2\lambda(\uu)}$ is called the conformal factor, which varies with position $\uu =\{u,v\}$ on spheroid. Because of the rotational symmetry $u=\phi$ and conformal factor depends only on parameter $v$. By equating two metrics we find that 
\begin{equation}\label{eq:conform}
e^{2\lambda(v)} = \frac{R^2\sin^2\theta}{a^2\sin^2 v}, \qquad\qquad\frac {d\theta}{\sin \theta}=\pm dv\,\sqrt{\cot^2 \!v+\frac{c^2}{a^2}}.
\end{equation}
By integrating both sides of the second equality we get
\begin{equation}\label{eq:gconf}
\log\bigg(\tan\frac\theta2\bigg)=g(v)\equiv-\sqrt{-1+\eta^2}\atan\Bigg(\frac{\sqrt{-1+\eta^2}\cot v}{\sqrt{\eta^2+\cot^2 v}}\Bigg)-\log\big(\cot v+\sqrt{\eta^2+\cot^2v}\big),
\end{equation}
where $\boxed{\eta=c/a}$. Substituting $\sin\theta=2\tan (\theta/2)/(1+\tan^2\theta/2)$ into Eq.~(\ref{eq:conform}) we find the analytical expression for the conformal factor 
\begin{equation}
e^{2\lambda(v)} = \frac {R^2}{a^2\sin^2\!v\,\cosh^2 \!g(v)}, 
\end{equation}
and the limit of interest 
\begin{equation}\label{eq:lambda}
\lim_{v\to0} \lambda(v) = \log\frac R a - \sqrt{-1+\eta^2} \atan\sqrt{-1+\eta^2}.
\end{equation}

\section{Free energy}

\subsection{Contribution from defects}

\begin{figure}[h]
\raisebox{47mm}{(a)\ }\includegraphics[height=5cm]{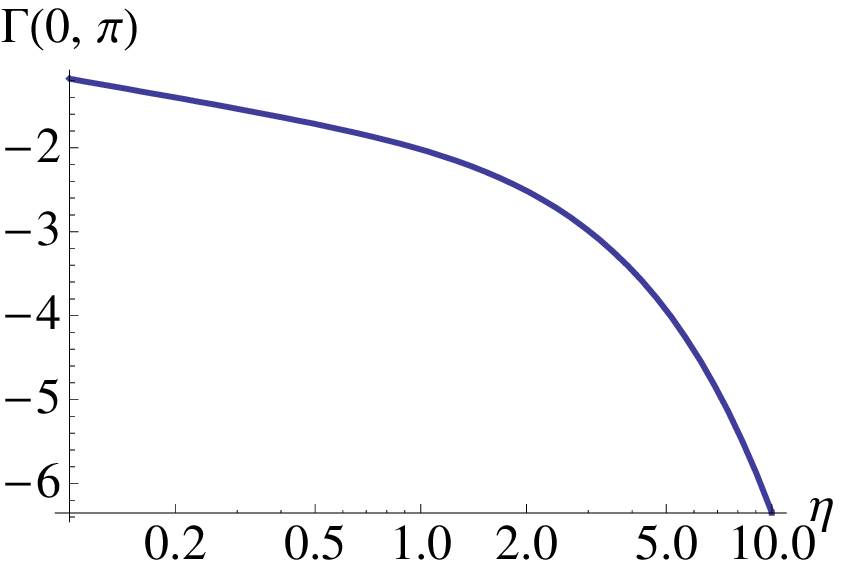} \hfil
\raisebox{47mm}{(b)}\includegraphics[height=5cm]{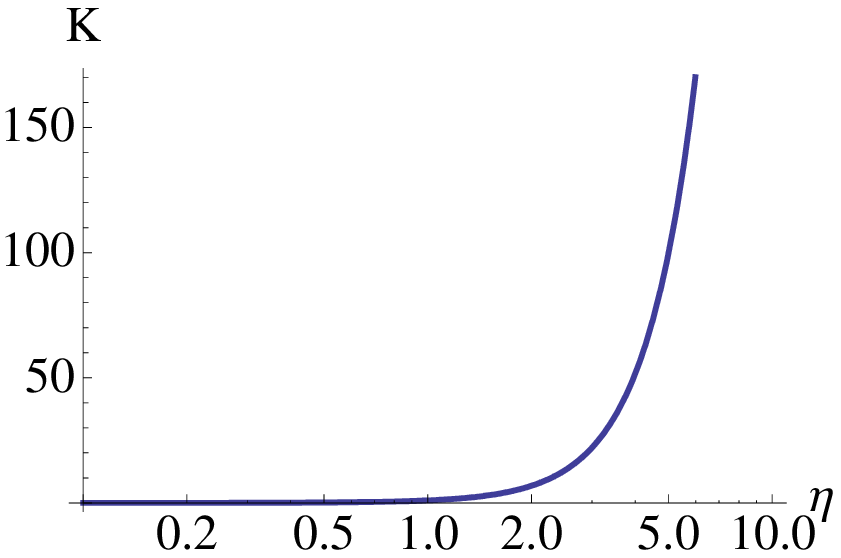}
\caption{(a) The Green function $\Gamma (0,\pi)$ given by Eq.~(\ref{eq:gamma}). For this plot we choose the equilibrium radius of the vesicle $R=100$~nm and the size of the core $d=3.5$~\AA as the equilibrium distance between thiophene cores. (b) The Gaussian curvature $K=c^2/a^4=\eta^2/a^2$ calculated at the poles, assuming the condition of constant surface (Eq.~\ref{eqSconst}).}
\label{figG}
\end{figure}

The pair Green function for topological defects on deformed sphere can be defined similar as for superfluids in reference~\cite{vitelli:2009}
\begin{equation}\label{eq:green}
\Gamma (\x_i,\x_j)=-\frac1{2\pi}\log\frac{{\cal D}_{ij}}d +\frac 1{4\pi} \big(\lambda(\x_i) +\lambda(\x_j) \big),
\end{equation}
where ${\cal D}_{ij}$ is the distance between two defects on the sphere (the chord between two points!) and $d$ is the core size of the defect\footnote{At this microscopic scale the continuum theory breaks, so $d$ may be thought as a cut-off radius.}. This representation of the Green function already includes both the interaction between defects (first term) and the position dependent self-energy of defect. In the case of two topological defects on sphere with topological charge $q_i=q_j=+1$, the minimum of $F_{\rm defect}=4\pi^2 K_A q_iq_j\Gamma (\x_i,\x_j)$ corresponds to the defects located at the north $N$ ($v=0$) and the south $S$ ($v=\pi$) poles (see Figure 1d). It gives ${\cal D}_{ij}=2R$, and together with Eqs.~(\ref{eq:lambda}) and (\ref{eq:green}) we find the Green function
\begin{equation}\label{eq:gamma}
\Gamma (0,\pi) = -\frac1{2\pi}\bigg(\log\frac{2a}d + \sqrt{-1+\eta^2} \atan\sqrt{-1+\eta^2}\bigg),
\end{equation}
and consequently an analytical expression for $F_{\rm defect}$. In Figure~\ref{figG} below we plot the Green function $\Gamma (0,\pi)$ and the Gaussian curvature $K$ as the function of $\eta$. As was expected, the Green function decreases with $\eta$, while the Gaussian curvature is increasing, resulting in the more negative $\Gamma (0,\pi)$ and thus $F_{\rm defect}$ for higher $K$, which is in agreement with a theory~\cite{vitelli:2004}.

\subsection{Bending energy}

The bending energy proposed by Helfrich in~\cite{helfrich:1973} is written as 
\begin{equation}\label{eq:bend}
F_{\rm bend} = 2k\iint \!dS\, H^2, 
\end{equation}
where $H$ is the mean curvature and the integral is over the surface $S$. For spheroids (Eq.~\ref{eqA:shape}) 
\begin{align}\label{eq:HS}
dS &= a\sin v \sqrt{\g}\,du\,dv,\\
H&=\frac12\Bigg(\frac{c/a}{\sqrt{\g}}+\frac{ac}{(\g)^{3/2}}\Bigg).
\end{align} 
After some calculations, the integral in Eq.~(\ref{eq:bend}) can be simplified to the following form 
\begin{equation}\label{eq:Fbend}
F_{\rm bend} = 2k\pi \eta^2\bigg\{\frac{\ath \d}\d+\frac2{\eta^2}+\frac1{3\eta^2}\bigg(1+\frac2{\eta^2}\bigg)\bigg\}.
\end{equation}
This energy term does not depend on the size of the vesicle only on the dimensionless parameter~$\eta$. However, the Green function in Eq.~(\ref{eq:gamma}) depends on the absolute value of the distance between the defects, therefore in all our calculations we assumed the condition of constant surface of spheroid
\begin{equation}\label{eqSconst}
S=2\pi a^2 \bigg(1+\eta^2 \frac{\ath \d}\d \bigg)=\const,
\end{equation}
yielding constant density of the molecules.

\section{Estimate of the ratio of $K_A/k$}

\begin{figure}[h]
\includegraphics[width=0.85\linewidth]{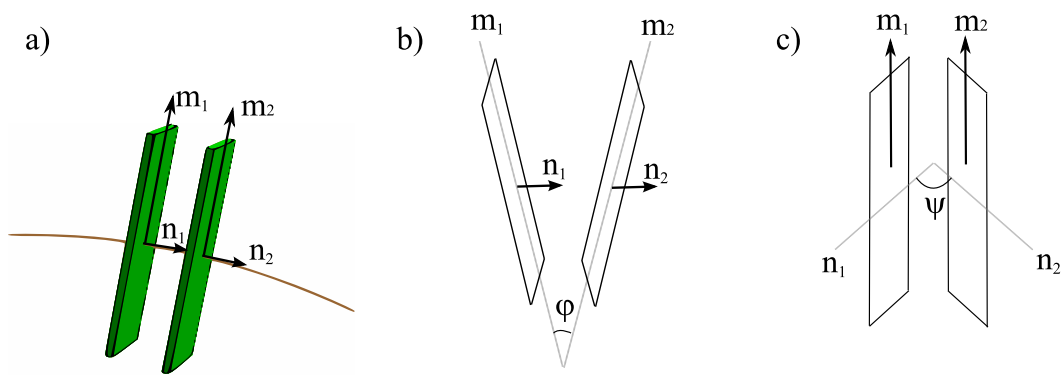}
\caption{(a) Flat plaquettes, described by unit vectors $n_i$ and $m_i$, which are normal and tangent vectors to the surface respectively. (b) The splay configuration of two plaquettes with the angle  $\phi$ defined as $\cos \phi= ({\bf m}_1,{\bf m}_2)$. (c) The rotation around the long axes of two plaquettes with the angle  $\psi$ defined as $\cos \psi= ({\bf n}_1,{\bf n}_2)$.}
\label{figK}
\end{figure}

We assume that for aromatic molecules this value is mainly determined by the long-ranged $\pi$-$\pi$ interactions between aromatic rings, and not by extrinsic effect (e.g. solvent) as discussed in~\cite{marzari:2004} for thiophene oligomers.  Therefore, we propose to estimate the ratio $K_A/k$ by using the results of quantum chemistry calculations for the binding energy of thiophene dimers with different relative orientations~\cite{tsuzuki:2002}. The basic idea is to associate the energy of the splay configuration of two $N$-thiophene molecules with the value of the bending rigidity $k$ (see Fig.~\ref{figK}b below), and the value of hexatic constant $K_A$ with the rotation around the long axes of the molecule (see Fig.~\ref{figK}c below). Then, based on the calculated interaction energy for different geometries of thiophene dimers (configurations $A$, $G$, $H$ of Figure~2, Table~2 of reference~\cite{tsuzuki:2002}) we find $k\propto E_{\rm total}(G)-E_{\rm total}(A)=-2.05-(-1.32)=-0.73$~kcal/mol $\approx 5\cdot 10^{-21}$~J and $K_A\propto E_{\rm total}(H)-E_{\rm total}(A)=-2.28-(-1.32)=-0.96$~kcal/mol, yielding $K_A/k\approx 1.3$. 

\bibliography{references}
\bibliographystyle{nar}



\end{document}